\documentclass[structabstract]{aa}  
\usepackage{graphicx}
\usepackage{amsfonts}
\usepackage{enumerate}
\usepackage{amssymb,amsmath}
\begin{document}
   \title{NoSOCS in SDSS III -- The interplay between galaxy evolution and the dynamical state of galaxy clusters}

\author{A.L.B. Ribeiro\inst{1}, P.A.A. Lopes\inst{2} \and S.B. Rembold\inst{1,3}}
        

   \offprints{A.L.B. Ribeiro}

   \institute{Laborat\'orio de Astrof\'{\i}sica Te\'orica e Observacional\\ 
              Departamento de Ci\^encias Exatas
e Tecnol\'ogicas\\ Universidade Estadual de Santa Cruz -- 45650-000, Ilh\'eus-BA, Brazil\\
		\email{albr@uesc.br}
              \and Observat\'orio do Valongo, Universidade Federal do Rio de Janeiro -- 20080-090, Rio de Janeiro-RJ, Brazil\\
\email{paal05@gmail.com}
\and Universidade Federal de Santa Maria -- 97105-900, Santa Maria-RS, Brazil}

   \date{Received xxxx; accepted xxxx}

 
  \abstract
  {We investigate relations between the color and luminosity distributions of cluster galaxies and the evolutionary state of their host clusters.}
{Our aim is to explore some aspects of cluster galaxy evolution and the dynamical state
of clusters as two sides of the same process.} 
{We used 10,721 member galaxies of 183 clusters extracted from the Sloan Digital
Sky Survey using a list of NoSOCS and CIRS targets. First, we classified the clusters into two categories, Gaussian and non-Gaussian, according to their velocity distribution measurements, which we used as an indicator of their dynamical state. We then used objective criteria to split up galaxies according to their luminosities, colors, and photometric mean stellar age. This information was then used to evaluate how
galaxies evolve in their host clusters.}
{Meaningful color gradients, i.e., the fraction of red galaxies as a function of radius from the center, are observed for both the Gaussian velocity distribution and the non-Gaussian velocity distribution cluster subsamples, which suggests that member galaxy colors change on a shorter timescale than the time needed for the cluster to reach dynamical equilibrium. We also found that larger portions of fainter red galaxies are found, on average, in
smaller radii. The luminosity function in Gaussian clusters has a brighter characteristic absolute magnitude and a steeper faint-end slope than it does in the non-Gaussian velocity distribution
clusters.} {Our findings suggest that cluster galaxies experience intense color 
evolution before virialization, while the formation
of faint galaxies through dynamical interactions probably
takes place on a longer timescale, possibly longer than the virialization time.}

\authorrunning{The interplay between galaxy and cluster evolution}
   \maketitle
%

\section{Introduction}

Galaxy ecology is an attempt to understand the environmental factors 
that affect galaxy formation and evolution (e.g. Balogh et al. 2004, Wetzel et al. 2012). 
There is strong evidence for
a relationship of galaxy properties and the physical conditions surrounding them
(e.g. Martinez et al. 2008).
At least three important pieces of information tell us that the environment affects
galaxy evolution in clusters. First, the morphology-density relation shows that galaxy morphologies
depend on the density (or cluster-centric radius) -- the
spiral fraction is larger at cluster outskirts than at the center (e.g. Dressler 1980; Whitmore, Gilmore \& Jones 1993).
Furthermore, the fraction of star-forming galaxies decreases with the increasing
density of the cluster environment (e.g. Balogh et al. 2004; Kauffmann et al. 2004).
Finally, there is a correlation between colors and environment -- the fraction of
red galaxies decreases toward the outer part of clusters 
(e.g. Balogh et al. 2000; De Propis et al. 2004).
Although these relations are connected, they correspond to
independent galaxy observables, and we can understand them as distinct
galaxy evolution indicators. Behind them, several physical mechanisms
may be present: tidal stripping, mergers, cannibalism,
strangulation, {\it ram} pressure stripping, galaxy harassment, etc. (see e.g. Mo, van den Bosh \& White 2010 for a good
description of all these transformation processes). 

In particular, the color-radius relation may be crucial for exploring the
environmental effects on galaxies. Because color evolution is mainly completed
at low redshifts (e.g. Goto et al. 2004), it would be interesting to relate the red galaxy
fraction to the dynamical state of their host clusters.
To probe the color-radius relation one needs to consider 
an important aspect, that is,  the very nature of the 
red galaxies, that is whether they are real
passive objects with no (or little) star formation currently
taking place (e.g. Strateva et al. 2001; Baldry et al. 2004).
Usually, it is assumed that all red galaxies are passive, but this 
neglects the possible contribution of dusty galaxies to the red sequence, which can be significant in an intermediate environment and for low-to-intermediate stellar masses (e.g. S\'anchez-Bl\'asquez et al. 2009).

An additional ingredient for galaxy ecology studies
is the role of low-luminosity galaxies. In the hierarchical scenario,
they have formed from small peaks in the initial power spectrum (e.g. White \& Rees 1978). 
If this is the case, they are expected to be the building blocks of the whole formation process (e.g. Evstgneeva et al. 2004).
We would find them in great numbers, and they would be less clustered than bright galaxies.
On the other hand, some of them may also have formed from tidal debris from galaxy
mergers or from galaxies that underwent {\it ram} pressure stripping as they
fell toward the center of clusters (e.g. Conselice et al. 2002) and, therefore, 
some of them would be expected at cluster centers (e.g. Trentham 1998, Carrasco et al. 2006). 
Hence, to determine the spatial
distribution and fraction of low-luminosity galaxies in clusters is important for 
understanding different aspects of galaxy formation and evolution models.

In this work, we present a study of the relation between colors and luminosities of cluster galaxies and
the dynamical state of the clusters.
Our aim is to describe the interplay between galaxy properties and  galaxy cluster evolution.
For this purpose, 
we used 10,721 member galaxies of 183 clusters extracted from the Sloan Digital
Sky Survey (SDSS) from a list of NoSOCS and CIRS targets. First, we 
applied statistical tests to classify the clusters into two categories, Gaussian and non-Gaussian, according to their velocity distribution, which we took as an indicator of their dynamical state. 
This was made irrespective of the triaxiality of the cluster halos. Even knowing that the orientation of the velocity ellipsoid is correlated with the large-scale structure and the 
anisotropic nature of infall into clusters (e.g. White et al. 2010), we assumed that these
effects do not severely influence the statistical approach used in this work. 

After classifying clusters
into Gaussian and non-Gaussian according to their velocity distribution, we
used objective criteria to split up the galaxies according to their luminosities, colors, and photometric mean stellar age. This information was finally used to evaluate how
galaxies evolve in their host clusters.

The paper is organized as follows: in Section 2 we describe our data; in Section
3 we present the methods employed to assess the dynamical state of the clusters and to separate
the galaxies according to 
color and luminosity; in Section 4 we analyze our results, in Section 5 we discuss
our findings,  and in Section 6 we present our conclusion.

\section{Data}

This work is based on the supplemental version of the Northern Sky 
Optical Cluster Survey (NoSOCS, Lopes 2003; Lopes et al. 2004). This
supplemental version of NoSOCS goes deeper ($r=21$ and $z \sim 0.5$), but covers a 
smaller region ($\sim 2,700$ square degrees) than the main NoSOCS catalog 
(Gal et al. 2003, 2009). NoSOCS was created from the digitized version of 
the Second Palomar Observatory Sky Survey (DPOSS/POSS-II, Djorgovski et 
al. 2003). The photometric calibration and object classification for DPOSS 
are described in Gal et al. (2004) and Odewahn et al. (2004), respectively.
In the first paper of this series (Lopes et al. 2009a) a subsample of
7,414 systems from the NoSOCS supplemental was extracted from the 
Sloan Digital Sky Survey (SDSS), data release 5 (DR5).

This sample was reduced to 179  low-redshift clusters ($z \le 0.1$) 
with enough spectra in SDSS (at least three galaxies within 0.50 h$^{-1}$ Mpc) 
for spectroscopic redshift determination using the gap technique 
(Katgert et al. 1996; Lopes 2007). This technique separates 
groups after the identification of gaps in the redshift distribution larger than a 
given value. To select members and exclude interlopers, the shifting gapper
technique (Fadda et al. 1996) was applied to all galaxies with spectra available 
within a maximum aperture of 2.50 h$^{-1}$ Mpc. This method works through the 
application of the gap technique in radial bins from the cluster center. The bin size 
is 0.42 h$^{-1}$ Mpc (0.60 Mpc for h = 0.7) or larger to force the selection of at least 
15 galaxies (consistent with Fadda et al. 1996). Galaxies not associated with the
main body of the cluster are eliminated. After removing the interlopers, the final sample
comprises 127 clusters (out of 179) with at least ten member galaxies within 
2.50 h$^{-1}$ Mpc. 

The line-of-sight velocity dispersion ($\sigma_P$) for these clusters was estimated 
and then a virial analysis was performed. The latter is analogous to the procedure 
described in Girardi et al. (1998), Popesso et al. (2005, 2007), and Biviano et al. (2006).
First, the projected, virial radius ($R_{PV}$) is derived and a first estimate of
the virial mass is obtained (using equation 5 of Girardi et al. 1998). The surface 
pressure correction is applied to the mass estimate and a Navarro et al. (1997) 
profile is assumed to obtain estimates of $R_{500}$, $R_{200}$, 
$M_{500}$ and $M_{200}$.

This low-redshift sample was complemented with 56
more massive systems from the cluster infall regions in the SDSS (CIRS) 
sample (Rines \& Diaferio 2006). CIRS is a collection of $z \le 0.1$ X-ray-selected 
clusters overlapping the SDSS DR4 footprint. In this redshift range, 
our NoSOCS sample therefore comprises only relatively poor systems. The same cluster 
parameters as listed above were determined for these 56 CIRS clusters. 
Hence, the combined NoSOCS plus CIRS sample comprises 183 clusters.
The NoSOCS clusters have velocity dispersion estimates of $100 <
\sigma_P < 700$ km/s, while the CIRS systems cover the range 
$200 < \sigma_P < 900$ km/s (only 23\% of CIRS objects have 
$\sigma_P < 400$ km/s). Because from the original CIRS member selection, velocity
dispersion and mass estimates are the product of the caustic
technique (Rines \& Diaferio 2006), CIRS was also used for comparing 
these properties and for the cluster scaling relations (Lopes et al. 2009ab).
For all the objects in our final we also estimated the sample optical luminosity ($L_{opt}$), the richness 
(N$_{gals}$), and the X-ray luminosity ($L_X$) ($L_X$ is obtained with 
ROSAT All Sky Survey data; see Lopes et al. 2009ab).

The centroid of each NoSOCS cluster is a luminosity, weighted estimate, 
which correlates well with the X-ray peak 
(see Lopes et al. 2006, and more details in 
Lopes et al. 2004). Lopes et al. 2009b, showed that the scaling 
relations based only on NoSOCS or CIRS objects agree, although in 
different mass ranges, which also indicates that the centroid (optical or X-ray) 
does not create a bias in the cluster parameters (richness, $L_{opt}$, 
$\sigma_P$, $R_{200}$, $M_{200}$), at least for our sample and analysis. Previous
works have also shown a good correlation between optical and X-ray centroids 
(Adami et al. 1998; Dai et al. 2006; Man \& Ebeling 2012). Note that Man \& Ebeling 
(2012) selected disturbed clusters in their search for objects with large offsets between 
the X-ray emission and the brightest cluster galaxy (BCG). Recently, centroid offsets 
have also been investigated using the Sunyaev-Zeldovich Effect (SZE, Sehgal et al. 2013) 
and gravitational lensing (Zitrin et al. 2012). Investigating the offsets between the
dark matter (DM) projected center and the BCG, Zitrin et al. (2012) found that some of 
the offsets are caused by misidentifications of the BCG.

The redshift limit of the sample ($z = 0.1$) is 
due to incompleteness in the SDSS spectroscopic survey for higher 
redshifts, where galaxies fainter than $M^* + 1$ are lacked, which biases 
the dynamical analysis (see the discussion in section 4.3 of 
Lopes et al. 2009a). We considered the $M^* = -20.94$ value from 
Popesso et al. (2006), converted to our cosmology.\footnote{
All quantities with cosmological dependence are computed in the concordance context,
defined by $\Omega_m$ = 0.3, $\Omega_\lambda$ = 0.7, and $H_0 = 100~h~{\rm km~s^{-1}Mpc^{-1}}$, with $h=0.7$.} Our galaxy sample consists of 10,721 members from these 183 clusters at
$z \le 0.1$.

To compute the absolute magnitudes of each galaxy (in $ugriz$ bands) we 
considered the following formula: $M_x = m_x - DM - k_{corr} - Qz$ ($x$ is one 
of the five SDSS bands we considered), where $DM$ is the distance modulus 
(considering the redshift of each galaxy), $k_{corr}$ is the $k-$correction
and $Qz$ ($Q = -1.4$; Yee \& Lopez-Cruz 1999) is a mild evolutionary 
correction applied to the magnitudes. The 
$k-$corrections are obtained directly from the SDSS database for every 
object in each band. Rest-frame colors are also derived for all objects.
More details regarding the sample (value of $M^*$, centroid 
determination, member selection, virial analysis, $L_X$, $L_{opt}$, and 
N$_{gals}$ estimates) can be found in Lopes et al. (2009ab).

\begin{figure}
\centering
    \includegraphics[width=84mm]{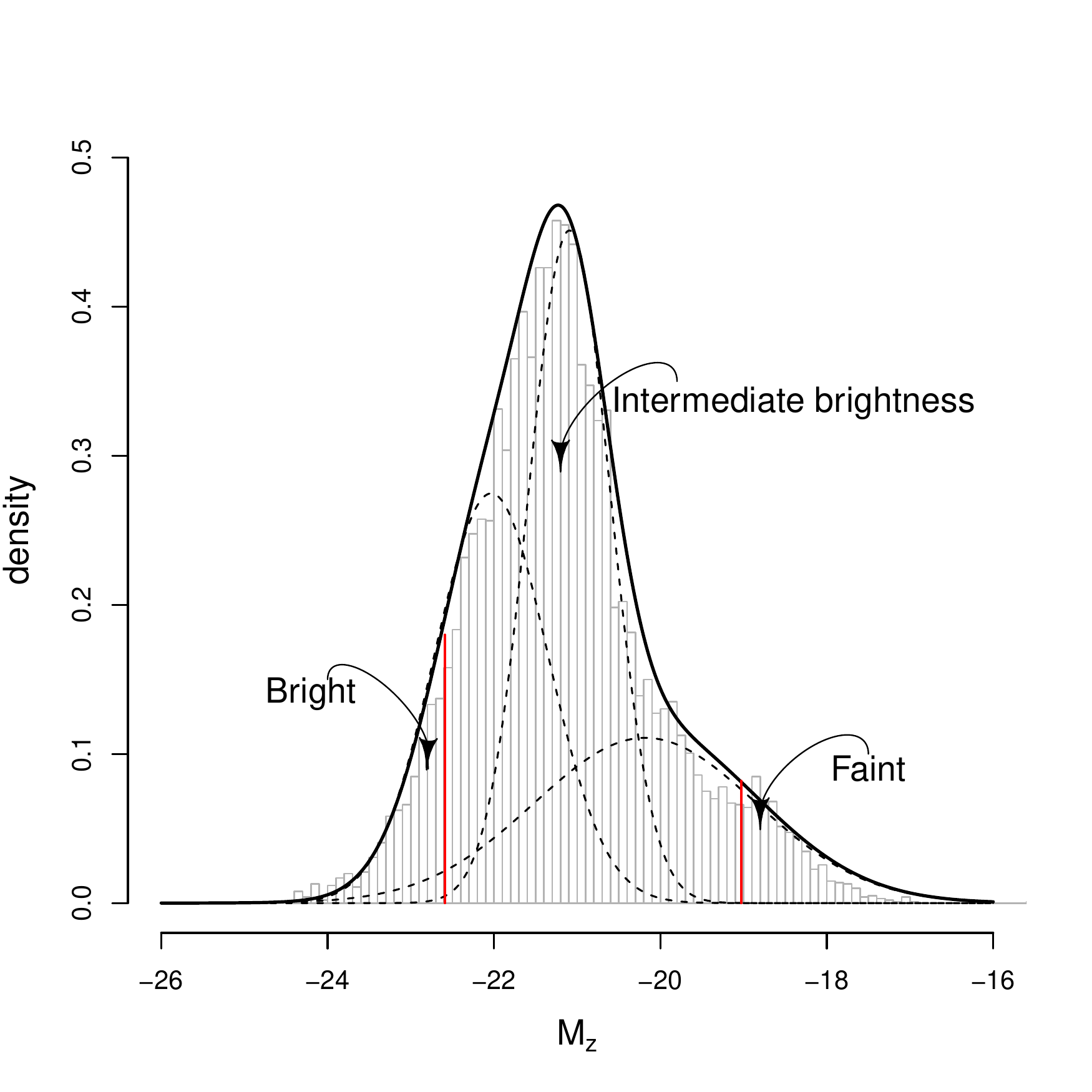}
\caption{Gaussian mixture model for the $M_z$ distribution. Dashed lines depict the individual components. 
The vertical red lines mark the intervals for
these modes: $M_z < -22.59$ for the bright population, $M_z > -19.03$ for faint population, and intermediate values of $M_z$ for the intermediate brightness population of galaxies. }
\end{figure}

\section{Methods}
\subsection{Gaussian and non-Gaussian composite clusters}

Our aim is to find a connection between galaxy evolution and the dynamical 
state of clusters. We started by defining two classes for the galaxy environments according to their velocity distributions: Gaussian velocity distribution and non-Gaussian velocity distribution clusters. Hereafter we refer to them as Gaussian and non-Gaussian clusters for simplicity, which represent the relaxed and less or unrelaxed samples. Although the theoretical  line-of-sight velocity distribution expected for galaxy clusters
is not exactly Gaussian (e.g. Merrit 1987),
phenomenological evidence has suggested for a long time that normality can
be assumed for clusters in dynamical equilibrium (e.g. Yahil \& Vidal 1977). Hence, distinguishing
galaxy clusters according to their velocity distributions may be an objective way to assess 
their dynamics through the simple use of statistical tests (see e.g. Hou et al. 2009, Ribeiro et al. 2010, 2011). 

Differently from our previous works that were only based on the Anderson-Darling (AD) test (usually appropriate for
small samples -- see Hou et al. 2009), we now used three more normality tests:
the Kolmogorov-Smirnov, the Shapiro-Wilk, and the Robust Jarque-Bera test (see Thode 2002 for
a comprehensive review on normality tests). We also used the dip test to probe the modality
of the velocity distributions: recent works have indicated that multimodal velocity distributions 
may be very common in galaxy clusters.
They probably even correspond to the
 main cause of non-normality in the velocity distribution in clusters
(e.g. Ribeiro et al. 2011, Hou et al. 2012, Einasto et al. 2012ab, Krause et al. 2013). Therefore, the dip method  
help us to better estimate which the intrinsically Gaussian velocity distributed clusters are and which are not.
The dip method is based on the cumulative distribution of the variable of interest (Hartigan \& Hartigan 1985). The dip statistic is the maximum distance between the cumulative input distribution and the best-fitting unimodal distribution. In some sense, this test is similar to the Kolmogorov-Smirnov test, but the dip test specifically searches for a flat step in the cumulative distribution function, which corresponds to a ``$dip$" in the histogram representation. The dip test has the benefit of being insensitive to the assumption of Gaussianity and is therefore a true test of modality (e.g. Pinkney et al 1996; Muratov \& Gnedin 2010).

We defined a non-Gaussian cluster when the dip test rejected unimodality
and at least one of the remaining tests rejected normality, taking member galaxies out to 2$R_{200}$.
For ten clusters with fewer than
eight members within this limit, we obtained no reliable results. These clusters were removed from
the sample. For the remaining clusters,
we found 146 Gaussian (84\%) and 27 non-Gaussian (16\%) clusters
with a total of 9,113 galaxies. This proportion is
approximately consistent with that found by Ribeiro et al. (2010).

We then built two stacked clusters, Gaussian and non-Gaussian, containing
6,478 and 2,635 galaxies, respectively. The distances of the 
galaxies in these composite groups are normalized to the distances to the group centers by $R_{200}$ and their velocities refer to the cluster median velocities and are scaled by the cluster velocity dispersions

\begin{equation}
u_i={{v_i - \langle v \rangle_j}\over \sigma_j},
\end{equation}

\noindent where $i$ and $j$ are the galaxy and the cluster indices, respectively. The velocity 
dispersions of the composite clusters, $\sigma_u$, refer to the dimensionless quantity $u_i$
(see Ribeiro et al. 2010, 2011).

The virial properties of non-Gaussian clusters were corrected after an iteratively removing galaxies without which
the clusters are Gaussian. On average, 15\% of the galaxies in non-Gaussian clusters need to be
removed in this process.The corrected properties are just those the cluster would have if it consisted only
of galaxies consistent with the normal velocity distribution (see Perea et al. 1990; Ribeiro et al. 2011).
This correction allows one to approximately compare virial properties of Gaussian and non-Gaussian clusters. After
computing the corrected 
virial properties, we returned the removed galaxies to non-Gaussian clusters for the subsequent photometric
analysis of the work.

\begin{figure}
\centering
    \includegraphics[width=84mm]{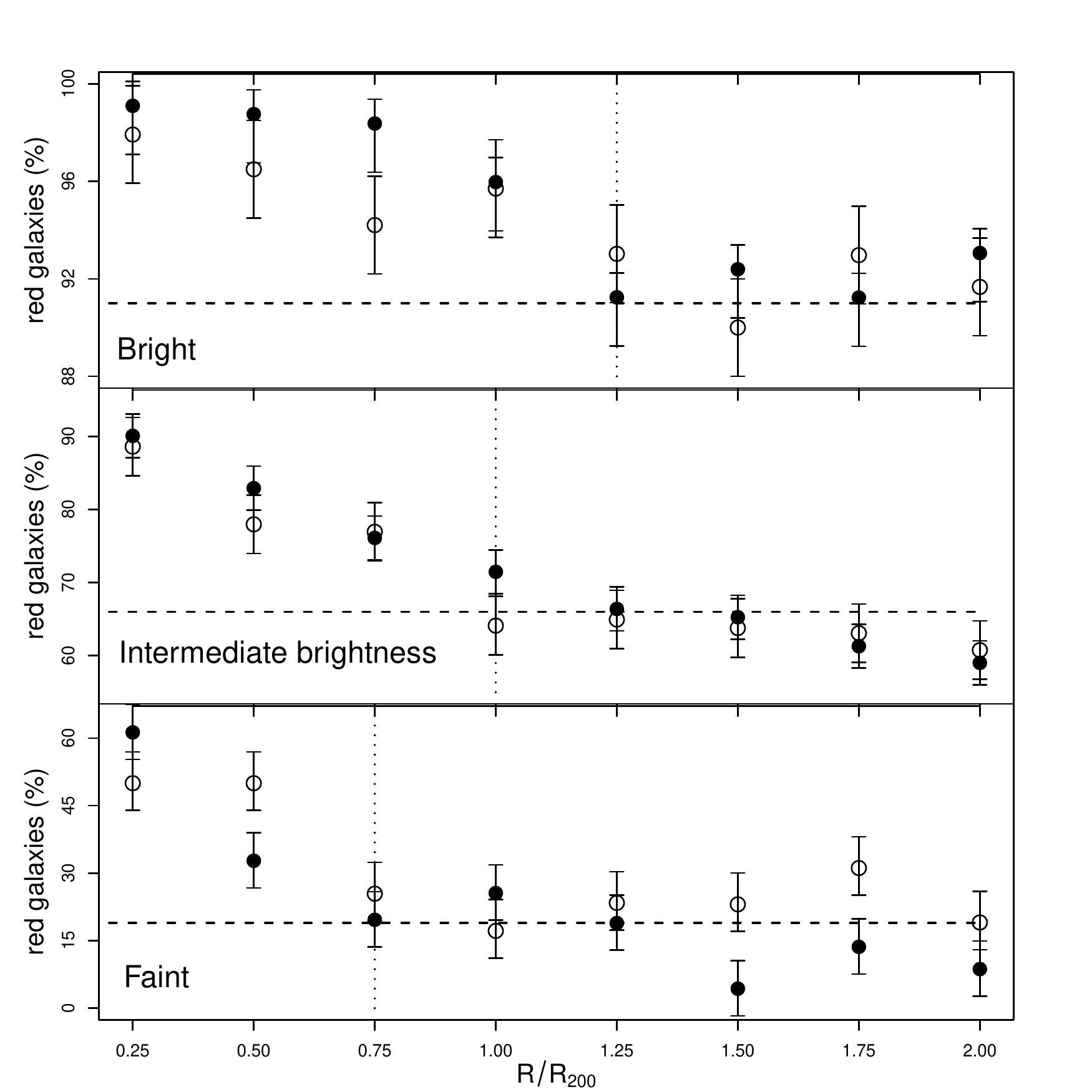}
\caption{Variation of the red galaxy fraction for the luminosity classes with
normalized clustercentric distance $R/R_{200}$ for
G (filled circles) and NG (open circles) clusters. Error-bars are obtained from a bootstrap technique with 1,000 resamplings.
The horizontal dashed lines indicate the RGF mean behaviour for the galaxy field. Vertical dotted lines indicate the stabilization radius for each component.}
\end{figure}

\subsection{Color and luminosity classification}

After setting the composite systems, we classified galaxies into blue or red objects.
Since the distribution of galaxies in the $g - r$ versus $u - g$ color-color diagram is strongly bimodal,
we simply classified galaxies 
above and below the separator developed by Strateva et al. (2001) as red and blue, respectively. 
Strateva et al. (2001) found that the blue galaxies are dominated by late types (spirals), while the red galaxies are dominated by early types (ellipticals), and that the $u - r > 2.22$ color-selection criterion for early types results in a completeness of 98\%. 

We also classified galaxies according to their 
luminosities. For this we fitted the  absolute magnitude distribution in the $z-$band ($M_z$) 
as a Gaussian mixture with $n$
components. We used the $z$-band absolute magnitude as the closest approximation to the galaxy mass
among the five SDSS color bands. It is estimated that the variation in
the stellar mass-to-light ratio is only a factor of $\sim$3 for galaxies in 
different ranges of the $z$-band absolute magnitude (Kauffmann
et al. 2003). In Figure 1, we see the $M_z$ histogram and 
the Gaussian mixture, fitted
after using a Bayesian regularization for normal mixture estimation  to find the
mixture parameters (Fraley \& Raftery 2007). The best fit is found for a mixture of
three normal distributions. In Figure 1, we used red vertical lines to mark the intervals for
these populations, corresponding to the first
quantile of the brightest component and the third quantile of the faintest component. 
This choice was made to avoid overlaps of the low- and high-luminosity ends
with the central and dominant luminosity population.

We considered a loose definition for each luminosity class, calling 
objects with $M_z < -22.59$ the bright population (corresponding to 15\% of all
galaxies), those with $M_z > -19.03$ as the faint population (7\% of all galaxies), 
and objects with intermediate 
values of $M_z$ correspond to the intermediate-brightness population of galaxies (78\% of all galaxies).
Finally, we find that 78\% of the bright, 72\% of the intermediate-brightness, and 56\% of the faint galaxies 
are in the stacked Gaussian cluster.

\section{Analysis}

\subsection{The red galaxy fraction}

To study the connection between galaxy evolution and the dynamical state of 
clusters, we looked for differences in the galaxy subsamples according to the
cluster dynamical classification, i.e., a Gaussian or non-Gaussian cluster. First, we used the 
red galaxy fraction (RGF) as an indicator of galaxy evolution
in the stacked clusters. We found that red galaxies correspond to 74\% of all galaxies.
In the stacked Gaussian cluster red galaxies correspond to 76\% of the total, 
while we found a somewhat lower fraction of red galaxies in the stacked non-Gaussian cluster, 70\%.

The mean RGF for a sample of field galaxies 
was estimated for each luminosity population and used for comparison.
The field sample, containing  68,375 objects, was constructed as follows. From the whole SDSS DR7 data set,
we selected galaxies that are not associated to a group or cluster,
considering the cluster catalog from Gal et al. (2009). To be conservative,
we defined a galaxy to belong to the field if it is not found within 4.0 Mpc
and does not have a redshift offset smaller than 0.06 of any cluster from
Gal et al. (2009).  Although this definition may include void galaxies in the sample,
this should not be a major problem, given that several studies have shown that 
color bimodality is remarkably similar between 
void and field galaxies (e.g. Patiri et al. 2006; Hoyle et al. 2012).
Finally, we divided the galaxies into the luminosity populations defined in Section 3.2, finding
4\%, 92\%, and 4\% of field galaxies in the bright, intermediate and faint populations, respectively.

In Figure 2, we observe
the emergence of the color-radius relation in all cases. The 
RGF decreases from the center toward the outskirts, with a slight 
recovering and/or stabilization to the outer radii, when
the behavior is approaching the field. 
 In each case we found a significant relationship between the RGF and the clustercentric distance,
with a Pearson correlation coefficient $\gtrsim 0.70$ at the 95\% confidence level.
This result is expected for Gaussian clusters,
which are supposed to be virialized. But we found the same behavior for non-Gaussian clusters.
This indicates that the color-radius (density) relation is already
set for non-Gaussian clusters, suggesting that a strong galaxy evolution happens
before the dynamical equilibrium is reached. 
 Since non-Gaussian clusters (as defined in this work) should be multi-modal,
our findings suggest that galaxies may have been pre-processed in the individual modes 
that constitute the clusters (i.e., 
the subunits we found in their velocity distribution -- see Ribeiro et al. 2011), a scenario consistent 
with that of Zabludoff \& Mulchaey (1998), where galaxies were pre-processed in
group environments before accretion into large clusters.

We also note in Figure 2 important differences with respect to the
point at which the RGF crosses the line that depicts the mean fraction of red galaxies in the field.
Our stacked analysis shows that larger portions of fainter red galaxies are found, on average, in
smaller radii.
 This result is probably related to the fact that the fraction of red galaxies heavily depends
on their own stellar mass (e.g. Baldry et al. 2006; Haines et al. 2006).
The most massive galaxies have resided within group-sized halos for a longer time
(e.g. MacGee et al. 2008), which could lead to a less steep trend for the RGF.

\subsection{The stellar population properties}

 To investigate the stellar 
population properties of galaxies in Gaussian/non-Gaussian groups, 
we used the stellar population synthesis
code STARLIGHT (Cid Fernandes et al. 2005; Asari et al. 2007).
This code fits the observed spectrum with a linear combination of a 
number of template spectra with known properties. We built a
template sample with single stellar population spectra from Bruzual \&
Charlot (2003) models, using a Chabrier (2003) IMF and the Padova 1994
stellar evolution tracks. Our template sample is composed of 45
spectra with three different metallicities ($Z=0.004$, $Z=0.02$, or
$Z=0.05$) and 15 ages ranging from 1\,Myr to 13\,Gyr. Prior to the
fits, we shifted the SDSS galaxy spectra to the restframe and
linearly resampled them to 1\,\AA~ (the spectra are expressed with
fixed resolution in logarithmic, not linear, scale). Our results are expressed
in terms of two parameters proposed by Cid Fernandes et al. (2005). The first
is the photometric mean stellar age,

\begin{equation}
<\log{t}>=\sum_j x_j \log t_j,
\end{equation}

\noindent where $x_j$ is the fractional contribution to the galaxy
total flux of the template $j$, and $t_j$ is the age of that
component (irrespective of its metallicity). The second is a measure of the
photometric age dispersion,

\begin{equation}
\sigma (\log{t})=\sqrt{\sum_j x_j (\log t_j - <\log t>)^2},
\end{equation}

\noindent which is higher for galaxies with extended star formation
history and zero for a single burst
of age $t$. Notice that due to the small SDSS fibre size (3\,arcsec)
the spectra measure the central
region of each galaxy, so that the inferred stellar population
distribution is not representative of the whole galaxy. 
However, even if a synthesis result cannot be translated into a
direct meaning because of this light sampling bias, we can
nevertheless identify average differences of the galaxies in the full
sample.

\begin{table}
      \caption[]{Mean properties of galaxies in each
component identified in the distribution of the photometric mean stellar age.}
$$
         \begin{array}{p{0.25\linewidth}ccccc}
		\hline
            \noalign{\smallskip}
            Component & N & M_z & \log{t} & \sigma(\log{t}) & u-r   \\
            \noalign{\smallskip}
            \hline
            \noalign{\smallskip}
            Old & 6094 & -21.43  & 9.90 & 1.07 & 2.69\\
            Transition & 774 & -21.08 & 9.32 & 1.52 & 2.42\\
            Young      & 2561 & -20.70 & 8.27 & 1.58 & 1.86\\
            \noalign{\smallskip}
            \hline
\end{array}
$$
\end{table}

The distribution of the photometric mean stellar age 
corresponds to a mixture of three components,  identified after using the Bayesian regularization for normal mixture estimation (the same
procedure we used in Section 3.2 to study the luminosity distribution; see Fraley \& Raftery 2007).
The components present the following
mean parameters: $\log{t}=9.90\pm 0.13$ (for the old component), 
$\log{t}=8.27\pm 0.27$ (for the young component) and
$\log{t}=9.32\pm 0.75$ (for what we call here the transition component).
Some properties of these components are presented in Table 1. The columns are: (1) the number
$N$ of galaxies; (2) the mean absolute magnitude in the $z$ band; (3) the
mean age in Gyr; (3) the mean age dispersion; and (4) the mean $u-r$ color.
These properties indicate that the old component is redder, more luminous
and has the lowest photometric age dispersion, on average,
showing that this component hosts a more homogeneous stellar population.

Galaxies in the color-age diagram have a clear bimodal distribution in this plane
-- see Figure 3 --, as expected for a mixture of blue and red galaxies. The horizontal green dashed line depicts the color separator
developed by Strateva et al. (2001), the vertical green dashed line depicts the
mean age of the transition component. We used these lines to roughly classify galaxies into
passive ($u-r > 2.22$ and $\log{t} > 9.32$) and star-forming objects 
($u-r \leq 2.22$ and $\log{t} \leq 9.32$).
The blue and red 1-$\sigma$ ellipses encompass the peaks identified in the color-age space
with the MCLUST code, 
a contributed R package for multivariate normal mixture modeling and model-based clustering.
\footnote{R is a language and environment for statistical computing and graphics. It is a GNU project that is similar to the S language and environment and was developed at Bell Laboratories -- R Development Core Team (2011).}
It provides functions for parameter estimation via the expectation-maximization (EM) algorithm for normal mixture models with a variety of covariance structures (Fraley \& Raftery 2007).
The ellipses indicate the central regions of star-forming (blue) and passive (red) galaxies.
We found that passive galaxies correspond to 84\% of all red objects (5877 out of 7020).
The remaining red galaxies are probably composed of dusty, star-forming
galaxies (e.g. Popesso et al. 2007), and were classified in the transition component.
Finally, star-forming objects correspond to 93\% of all blue objects (2231 out of 2409).

\begin{figure}
\centering
    \includegraphics[width=84mm]{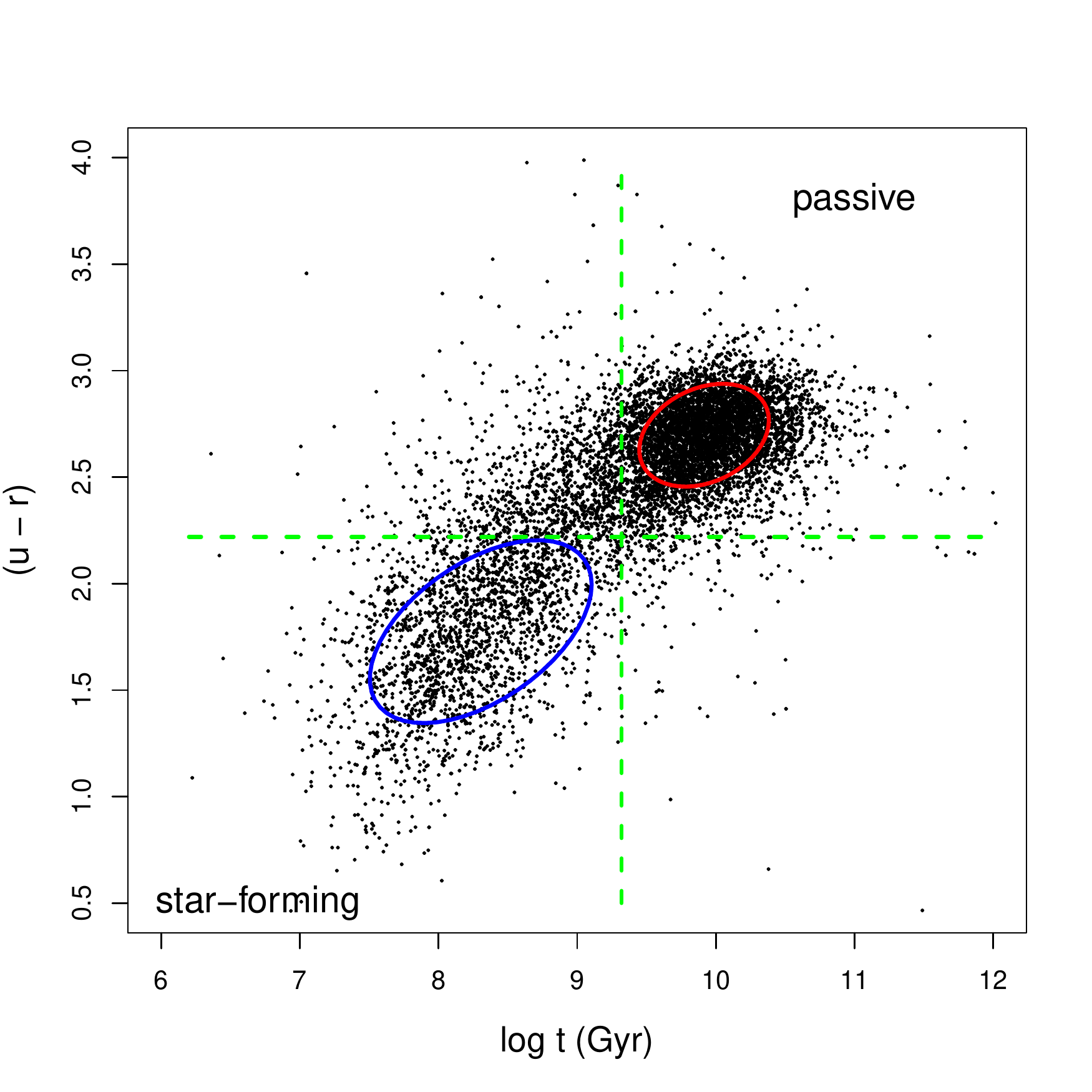}
\caption{Bimodal distribution of galaxies in the color-age diagram showing the 
1$\sigma$ ellipse around the star-forming (blue) and passive (red) peaks. The
horizontal and vertical green dashed lines indicate the red/blue and star-forming/passive
separations, respectively.}
\end{figure}


\subsection{Galaxy evolution in the stacked clusters}

We now study some properties of the stellar and luminosity populations in
the stacked clusters. 

\begin{figure}
\centering
    \includegraphics[width=84mm]{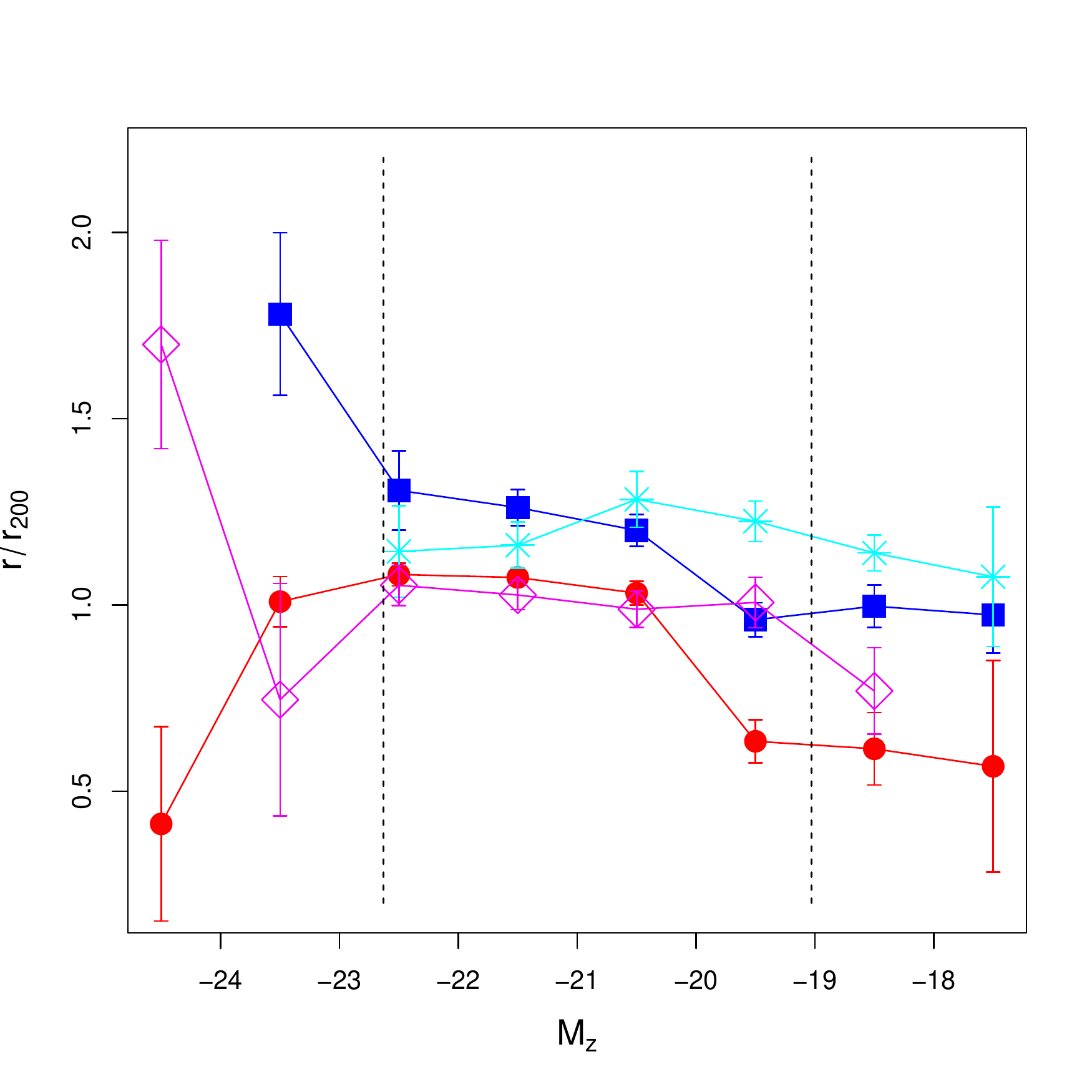}
\caption{Normalized clustercentric distance of galaxies as a function of their $z$-band
absolute magnitude. Red circles and magenta diamonds represent passive galaxies in Gaussian and
non-Gaussian clusters, respectively. Blue squares and cyan stars represent star-forming galaxies
in Gaussian and non-Gaussian clusters, respectively. The vertical dashed lines indicate the separation into
luminosity populations. 
}
\end{figure}

\begin{figure}
\centering
    \includegraphics[width=84mm]{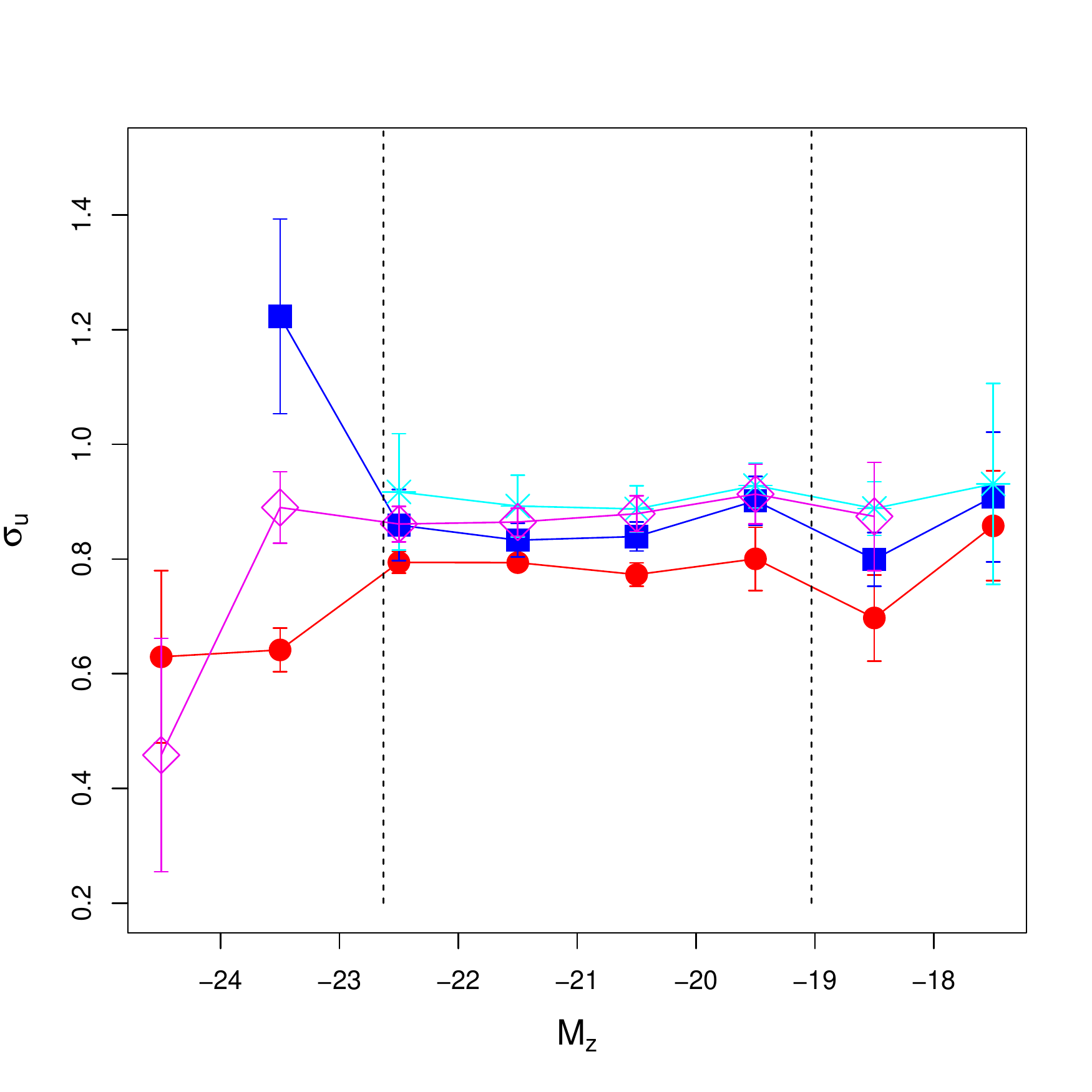}
\caption{Velocity dispersions of galaxies as a function of their $z$-band
absolute magnitude. Red circles and magenta diamonds represent passive galaxies in Gaussian and
non-Gaussian clusters, respectively. Blue squares and cyan stars represent star-forming galaxies
in Gaussian and non-Gaussian clusters. The vertical dashed lines indicate the separation into
luminosity populations.}
\end{figure}

\subsubsection{$M_z~\times~r/r_{200}$}

Figure 4 shows the normalized clustercentric distance of galaxies 
as a function of their $z$-band absolute magnitude. Passive galaxies in Gaussian clusters always
have smaller clustercentric distances than star-forming galaxies in both Gaussian and non-Gaussian clusters.
For galaxies in the range $-22.5<M_z<-20.5$, passive galaxies in Gaussian and non-Gaussian clusters
have the same behavior, while galaxies brighter than $M_z\approx -22.5$ and 
fainter than $M_z \approx -20.5$ are more centrally located in Gaussian than in non-Gaussian clusters.
The clustercentric distance of passive galaxies brighter than $M_z\approx -22.5$
is smaller toward brighter magnitudes in Gaussian clusters. 

Gaussian clusters seem to have a clear structure: passive galaxies in the
virialized region ($r<r_{200}$), and star-forming galaxies inhabiting the outer region ($r\gtrsim r_{200}$).
The absence of star-forming galaxies in the center indicates that the environment was able to
transform nearly all galaxies into red objects in the innermost regions. 
In fact, we see in Figure 4 that the clustercentric distance of star-forming galaxies 
brighter than $M_z\approx -19.5$ is systematically smaller toward fainter magnitudes in Gaussian clusters, 
suggesting that dynamical processes
may be providing both reddening and mass loss as these objects fall into the
potential well of the cluster.

The more central distribution of bright and faint passive galaxies in Gaussian clusters
can be intepreted as an indicator of dynamical evolution.
Luminosity segregation describes the tendency for more luminous/massive 
galaxies to be located near the cluster center (e.g. Rood et al. 1972; Dressler 1978). This effect is expected from
the dynamics of galaxies aggregating in the potential wells of clusters (e.g. Fusco-Femiano \& Menci 1998), and
thus clusters with luminosity segregation are considered to be more highly 
evolved than those without this feature.
We also see in Figure 4 that
faint passive galaxies show segregation with magnitude and position. If a fraction of these faint galaxies 
correspond to the result of galaxy-galaxy encounters in the cluster (e.g. Conselice 2002),
the presence of faint galaxies in the innermost regions of clusters could be
taken as an additional indicator of dynamical evolution.

Finally, we note that the spatial distribution of galaxies in non-Gaussian clusters 
 has some similarity to their counterparts in Gaussian ones, but with certain important
differences between them: (i) non-Gaussian clusters do not present luminosity segregation -- their
brightest galaxies are peripheral; (ii) star-forming galaxies in non-Gaussian clusters show no significant trend
with the clustercentric distance; (iii) there are fewer faint galaxies
close to the cluster center. These results may be indicating that, regarding the galaxy
population, non-Gaussian clusters are less evolved environments than Gaussian 
clusters.

\subsubsection{$M_z~\times~\sigma_u$}

The correlation between the velocity dispersion and the absolute
magnitude in the $z$-band of the galaxies is shown in Figure 5.
We call velocity dispersion the dispersion of
the galaxy velocities in each bin of absolute magnitude.
We confirmed the segregation between passive and star-forming galaxies
both in the Gaussian and non-Gaussian stacked clusters: the brightest objects
($M_z < -22.5$) show a lower velocity dispersion, which indicates that 
they had time to decrease their velocity
through dynamical interactions or friction. We also see that passive galaxies
in Gaussian clusters have a lower velocity dispersion over the entire luminosity range.
This result agrees well with some studies reporting that late-type galaxies
have a higher velocity dispersion than early types in nearby clusters
(e.g. Moss \& Dickens 1977; Sodr\'e et al. 1989; Adami et al. 1998, 2000). This is not
true for non-Gaussian clusters, where passive and star-forming galaxies
have approximately the same velocity dispersion in the range $-22.5< M_z < -18.5$.
We also note that the brightest star-forming galaxies in Gaussian clusters
have a high velocity dispersion and large clustercentric distance, which suggests
that these galaxies are falling into the clusters, possibly through
radial orbits (e.g. Biviano \& Katgert 2004). 

From Figures 4 and 5, we see that
bright passive galaxies in Gaussian clusters combine low-velocity dispersion with
small clustercentric distance. This kind of segregation with magnitude in
velocity and position has been observed in many clusters (e.g. Carlberg et al. 1997, Biviano \& Katgert 2004).
It can be understood with the model of Menci
\& Fusco-Femiano (1996), which is a solution of the
collisional Boltzmann-Liouville equation, and hence accounts for galaxy collisions and merging processes
in the cluster center. However, we also see in Figures 4 and 5 that
the brightest passive galaxies in non-Gaussian clusters
combine low velocity dispersion with large clustercentric distance, indicating
that these systems are less segregated than Gaussian clusters.
Indeed, it has been found that clusters present a gradation of luminosity segregation, from highly segregated to virtually not segregated, and that their different degrees could be a consequence of either the different initial conditions present at the onset of the two-body relaxation phase or the different cluster dynamical stages 
(e.g. Yepes et al. 1991, Ribeiro et al. 2010). We must also consider that observational evidence for 
luminosity segregation in distance may be weakened by subclustering
(Biviano et al. 1992).
According to our definition in Section 3.1, non-Gaussian clusters must be multi-modal,
which means that the centers of these systems are ill-defined and that
galaxy properties may refer more to the individual modes than to the
entire cluster. In Figure 4, we
see that both the brightest and the faintest passive galaxies in non-Gaussian clusters are
more peripheral than their counterparts in Gaussian clusters. We verified that
the ratio of the clustercentric distance of passive galaxies with respect to the mode center and
to the cluster center of non-Gaussian clusters are $\sim$0.41, $\sim$0.94, and $\sim$0.65
for the bright, intermediate, and faint populations, respectively. Thus, in the individual modes,
bright passive galaxies are the most centrally located, just like in Gaussian clusters,
and segregation in distance is recovered.

These findings agree with the idea of non-Gaussian clusters as a set of
subclusters that individually show a degree of dynamical evolution, regarding the
galaxy population. At the same time,
Gaussian clusters appear to be more evolved systems
with strong evidence of segregation,
and containing many faint galaxies in the central  regions of the clusters. If there is a sufficient number of
these faint galaxies, they may steepen  the faint end of the luminosity function of 
galaxies in clusters, as we show bellow.

\subsection{Luminosity function}

The results achieved hitherto suggest some 
important differences between Gaussian and non-Gaussian clusters. We can investigate 
this point in more detail through the luminosity function 
of galaxies in our cluster samples. 
We built the stacked central luminosity functions (LFs), inside 
a radius of 0.5\,$R_{200}$ from the center, using the
Colless (1989) cumulation method. This method consists of a summation of the number counts (corrected for the background contribution) in all magnitude bins of the individual LFs, normalized by the total number of galaxies within the magnitude completeness limit of each cluster ($M_r=-19$ for the NoSOCS sample). 
All photometric sources classified as galaxies that are located at
a maximum projected distance of 10 Mpc to the center 
of each cluster were retrieved from the SDSS DR7 database.
The absolute magnitude $M_r$ of each galaxy was obtained by $M_r=m_r-5\,\log\,D/10-K,$ where $K$ is the K-correction and $D$ is  the luminosity distance of the galaxy (in pc). All galaxies in the
direction of a given cluster were assigned the same value of $D$, determined from the cluster redshift; this is certainly not true for the background galaxies, but our results are insensitive to this because the background contribution can be accessed independently from the individual choice of $D$. The same K-correction was applied to all galaxies in a given cluster, corresponding to the K-correction of an elliptical galaxy at the cluster redshift. De Filippis et al. (2011) have also applied this methodology for the NoSOCS sample, and showed that 
the results are not much different when the individual K-corrections are 
determined for all galaxies in the photometric sample.
K-corrections were calculated as in Chilingarian, Melchior \& Zolotukhin (2010), using the $(g-r)$ colors of
elliptical galaxies of Fukugita, Shimasaku \& Ichikawa (1995).
The background contribution in the individual LFs was obtained in an area with 10$\times$10\,Mpc and outside a radius of 2.0\,$R_{200}$, where the contribution of cluster galaxies is expected to be very low. Local overdensities were excluded from the background estimation area whenever the local number counts exceeded 3$\sigma$ from the mean background, calculated in square cells 150\,kpc wide. We then rescaled the background number count in each magnitude bin to the area of the central circular region and subtracted it from the number counts in this region.

The resulting LFs for the Gaussian and non-Gaussian composite clusters are shown in Figure 6. 
This figure also shows the best-fit Schechter functions (artificially displaced for clarity) obtained by minimization of square residuals. 
For the non-Gaussian composite cluster, the fit resulted in $M_r^\ast=-21.31\pm 0.16$ and
$\alpha=-0.97\pm 0.06$. 
For the Gaussian composite cluster, a brighter characteristic magnitude $M_r^\ast=-21.90\pm 0.14$ and
a steeper faint-end with $\alpha=-1.14\pm 0.03$ were obtained. This result is consistent with the work of
Mart\'{\i}nez \& Zandivarez (2011), and shows that at least a fraction of low-luminosity galaxies
is correlated with the environment.

\begin{figure}
\centering
    \includegraphics[width=74mm]{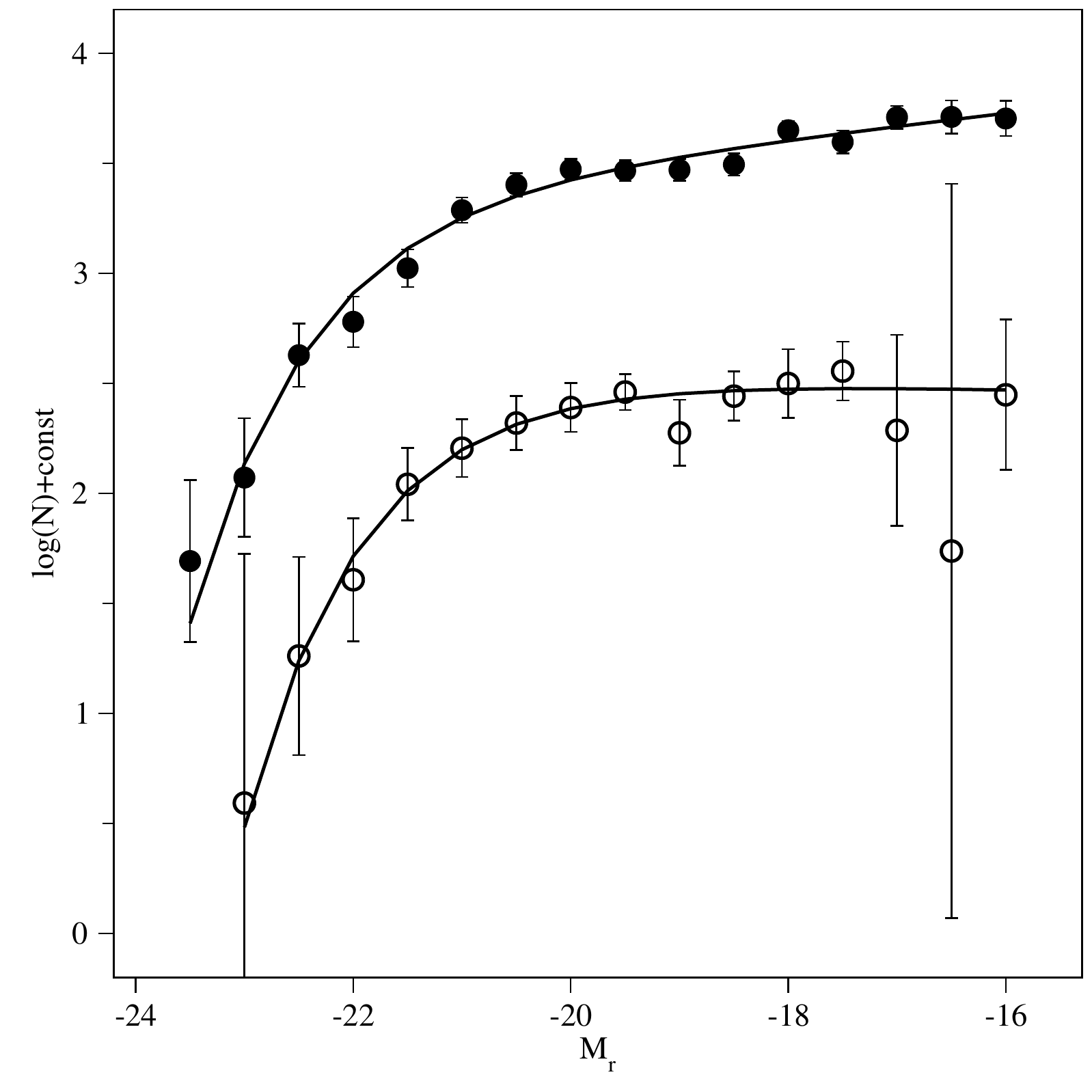}
\caption{Cumulative LFs of Gaussian (filled circles) and non-Gaussian (open circles) systems. The best-fit Schechter parameters are also shown.}
\end{figure}

\section{Discussion}

Our analysis gives some hints for understanding the interplay between 
galaxy evolution and the dynamical state of galaxy clusters.
We verified that Gaussian clusters present at least two important
properties expected for dynamically evolved environments:
(i) the color-radius relation, and (ii) the luminosity segregation in velocity and position.
We have to jointly consider these results to understand what 
they imply about the interplay between galaxy evolution
and the dynamical state of galaxy clusters. First, 
it is important to note that property (i) was also found in non-Gaussian clusters,
reinforcing the idea of a shorter timescale for color evolution than that 
of the dynamical evolution of the clusters. Goto et al. (2004) showed that
color evolution is almost completed by $z\sim0.2$ for a sample of galaxies in SDSS clusters, while
Zabludoff \& Mulchaey (1998) suggested that galaxies are pre-processed in group
environments before accretion into larger clusters. Thus, the color-radius relation is
expected for low redshifts, regardless of the dynamical state of the clusters, and hence 
it cannot be taken as an indicator of galaxy cluster evolution. 
On the other hand, there is an important difference between Gaussian and non-Gaussian
clusters with respect to the property (ii). While segregation in velocity
is well defined for passive galaxies in both stacked clusters, segregation in position is less well-defined
in non-Gaussian clusters due to the large clustercentric distance of their brightest passive galaxies, a
possible consequence of the multimodal nature of non-Gaussian clusters. Hence, 
considering properties (i) and (ii), our results indicate that non-Gaussian clusters are less segregated, 
because of subclustering, but with color evolution at an advanced stage in the
individual modes composing them.

At the same time, we found more faint galaxies in Gaussian clusters than in non-Gaussian clusters.
Faint galaxies in nearby clusters present a variety in their properties, with
significant differences in their spatial and kinematic distributions (e.g. Lisker et al. 2006, 
Toloba et al. 2009, Paudel et al. 2010). In particular, early-type dwarf galaxies play
an important role in understanding galaxy cluster evolution. Since these galaxies
have low mass and density, they are more susceptible to interactions with larger galaxies
in the cluster environment (e.g. Lisker et al. 2006). Possibly, the number of these galaxies results from 
the mixture of survival and destruction mechanisms in the cluster environment. Part of them could
have a primordial origin and start to be destroyed by tidal stripping as they fall into the clusters. 
They could also correspond to remnants of galaxy-galaxy encounters  (e.g. Dabringhausen \& Kroupa 2013). 
Alternatively, they could correspond to
primordial dwarf galaxies that were transported by galaxy groups from the outside to the inside of clusters
(e.g. S\'anchez-Janssen et al. 2008). 
This latter possibility  favors the idea of non-Gaussian
clusters as a crucial step in the evolutionary process of  galaxy clusters. They could be made of
interacting groups, each one transporting primordial faint galaxies to the center of future
more regular clusters. This possibility is consistent with the assumption that the tidal-stripping 
efficiency is proportional to the ratio of the cluster mass (M)
to the galaxy mass (m) (e.g. Kimm et al. 2009). If non-Gaussian
clusters correspond to assemblies of galaxy groups with smaller masses, then
we have lower M/m values and thus we should expect less tidal stripping in their environments.
That is, faint galaxies are expected to survive much longer in the modes of non-Gaussian clusters.

At the same time, if part of the faint passive galaxies was already present in non-Gaussian clusters,
this means that at least a fraction of these galaxies in Gaussian clusters are formed in the
cluster environment -- remember that Gaussian clusters present
an excess of faint galaxies. Indeed, Figure 4 shows that faint passive galaxies in Gaussian clusters
are more central than their counterparts in non-Gaussian clusters. 
The mean clustercentric distances are $\langle r/r_{200}\rangle^{\rm G}\simeq 0.62\pm 0.34$ and
$\langle r/r_{200}\rangle^{\rm NG}\simeq 0.91\pm 0.23$. The Kolmogorov-Smirnov test indicates
a significant difference between them
at the 95\% confidence level.\footnote{No significant difference was found for faint star-forming galaxies.} 
This difference in
the spatial distribution of these galaxies may be pointing to differences concerning their origin.
While slow-motion conditions are more common in low-mass groups, slow
galaxy–-galaxy interactions and mergers are likely to be rare in virialized clusters (e.g. Ostriker 1980).
Galaxies moving at high speeds and in random directions
in the center of massive galaxy clusters promote multiple 
rapid encounters between galaxies, known as galaxy harassment,
a possible mechanism underlying the production of faint passive galaxies in Gaussian clusters.
Aguerri \& Gonz\'alez-Garc\'{\i}a (2009) showed that fast galaxy-galaxy interactions
are efficient mechanisms to transform late-type galaxies into dwarf galaxies. Thus,
the presence of faint galaxies in the central region of Gaussian clusters (and their
lower incidence in non-Gaussian clusters) can be taken as a possible indicator of the
dynamical state of a galaxy cluster.

A more detailed study of the different types of faint galaxies in Gaussian and
non-Gaussian clusters is beyond 
the scope of the present work. We will investigate this point in a future contribution
based on a larger sample.

\section{Conclusion}

Understanding the process of galaxy cluster formation
and evolution is an important step toward refining the details of the structure growth and constraining
the evolution of galaxy populations in different environments. 
The analysis presented in this work is an attempt 
to link galaxy evolution with the dynamical state of clusters.
Our findings suggest a scenario where non-Gaussian clusters represent a previous stage in the
life of a galaxy cluster during which two or more galaxy groups
merge to eventually form Gaussian clusters. Non-Gaussian clusters
are less segregated because of subclustering, but with 
color evolution already at an advanced stage in the individual modes they are composed of.
The modes may also be carrying faint primordial galaxies into the central regions of larger clusters, while
the formation of faint galaxies through dynamical interactions 
probably takes place on a longer timescale, probably achieving the Gaussian phase.

\section*{Acknowledgments}
We acknowledge the referee for very helpful suggestions.
ALBR also thanks for the support of CNPq, grants 306870/2010-0 and 478753/2010-1. 
PAAL thanks for the support of FAPERJ, process 110.237/2010.


\begin{thebibliography}{99}
\bibitem[]{} Adami, C. et al., 1998, A\&A, 336, 63
\bibitem[]{} Adami, C., Holden, B.P., Castander, F.J., Mazure, A., Nichol, R.C. \& Ulmer, M.P.,
2000, A\&A, 362, 825
\bibitem[]{} Aguerri, J.A.L. \& Gonz\'alez-Garc\'{\i}a, A.C., 2009, A\& A, 494, 891
\bibitem[]{} Asari, N.V., Cid Fernandes, R., Stasinska, G., 
Torres-Papaqui, J.P., Mateus, A., Sodré, L., Jr., Schoenell, W. \& Gomes,
J. M., 2007, MNRAS 381, 263
\bibitem[]{} Bah\'e, Y.M., McCarthy, I.G., Crain, R.A. \& Theuns, T., 2012, MNRAS, 424, 1179
\bibitem[]{} Baldry, I.K. et al., 2004, ApJ, 600, 681
\bibitem[]{} Balogh, M., Morris, S.L., Yee, H.K.C., Carlberg \& R.G., Ellingson, E., 1999, ApJ, 527, 54
\bibitem[]{} Balogh, M., Navarro, J. \& Morris, S.L., 2000, ApJ, 540, 113
\bibitem[]{} Balogh, M., Baldry, I.K, Nichol, R., Miller, C., Bower, R. \& Glazebrook, K., 2004, ApJ, 615, L101
\bibitem[]{} Balogh, M. et al., 2004, MNRAS, 348, 1355
\bibitem[]{} Balogh, M. \& McGee, S., 2010, MNRAS, 402, L59
\bibitem[]{} Biviano A., Katgert P., 2004, A\& A, 424, 779
\bibitem[]{} Biviano A., Girardi, M., Giuricin, G., Mardirossian, F. \& Mezzetti, M., 1992, ApJ, 396, 35
\bibitem[]{} Biviano A., Katgert P., Thomas, T. \& Adami, C., 2002, A\& A, 387, 8
\bibitem[]{} Biviano A., Murante G., Borgani S.,
Diaferio A., Dolag K., Girardi M. 2006, A\&A, 456, 23
\bibitem[]{} Bruzual, G. \& Charlot, S., 2003, MNRAS, 344, 1000
\bibitem[]{} Carlberg, R.G., Yee, H.K.C., Ellingson, E. et al., 1997, ApJ, 476, L7
\bibitem[]{} Carrasco, E.R., Mendes de Oliveira, C. \& Infante, L., 2006, AJ, 132, 1796
\bibitem[]{} Cid Fernandes, R, Mateus, A., Sodr\'e, L., Jr., Stasinska, G. \& Gomes, J.
M., 2005, MNRAS 358, 363
\bibitem[]{} Chabrier, G. 2003, PASP, 115,763
\bibitem[]{} Chilingarian, I. V., Melchior, A.-L. \& Zolotukhin, I. Y., 2010, MNRAS, 405, 1409
\bibitem[]{} Colless, M., 1989, MNRAS 237, 799
\bibitem[]{} Conselice, C.J., 2002, ApJ, 573, L5
\bibitem[]{} Conselice, C.J., Gallagher, J.S. \& Wyse, R.F.G., 2002, AJ, 123, 2246
\bibitem[]{} Dabringhausen, J. \& Kroupa, P., 2013, MNRAS, 429, 1858
\bibitem[]{} Dai, X. et al., 2007, ApJ, 658, 917
\bibitem[]{} de Filippis, E., Paolillo, M., Longo, G., La Barbera, F., de Carvalho, R. R., Gal, R., 2011, MNRAS, 414, 2771
\bibitem[]{} De Propis, R. et al. 2004, MNRAS, 351, 125
\bibitem[]{} Djorgovski S.G., de Carvalho R.R., Gal R.R., Odewahn S.C., Mahabal A.A., Brunner R.J.,
Lopes P.A.A., Kohl Moreira J.L. 2003, Bulletin of the Astronomical Society of Brazil, 23, 197
\bibitem[]{} Dressler, A. 1978, ApJ, 226, 55
\bibitem[]{} Dressler, A., 1980, ApJ, 236, 351
\bibitem[]{} Drinkwater, M.J., Gregg, M. \& Colless, M., 2001, ApJ, 548, 139L
\bibitem[]{} Einasto, M., Vennik, J., Nurmi, P.,Tempel, E.,Ahvensalmi A., Tago, E.,Liivam\"agi, L. J.,  Saar, E.,  Hein\"m\"ki, P., Einasto, J.,  \& Martínez, V. J., 2012a, A\& A, 540, id.A123 
\bibitem[]{} Einasto, M., Liivam\"agi, L. J., Tempel, E., Saar, E., Vennik, J., Nurmi, P., Gramann, M., 
Einasto, J., Tago, E., Hein\"m\"ki, P., Ahvensalmi, A. \& Martínez, V. J., 2012b, A\& A, 542, id.A36
\bibitem[]{} Evstgneeva, E., de Carvalho, R.R., Ribeiro, A.L.B. \& Capelato, H.V., 2004, MNRAS, 349, 1052
\bibitem[]{} Fadda D., Girardi M., Giuricin G., et al., 1996, ApJ, 473, 670 
\bibitem[]{} Fasano, G. \& Franceschini, A., 1987, MNRAS, 225, 155
\bibitem[]{} Fraley, C. \& Raftery, A.E, 2007,  Journal of Classification 24, 155
\bibitem[]{} Fukugita, M., Shimasaku, K. \& Ichikawa, T., 1995, PASP, 107, 945
\bibitem[]{} Fusco-Femiano, R. \& Menci, N., 1998, ApJ, 498, 95
\bibitem[]{} Gal R.R., de Carvalho R.R., Lopes P.A.A., Djorgovski S.G.,
Brunner R.J., Mahabal A.A., Odewahn S.C. 2003, AJ, 125, 2064
\bibitem[]{} Gal R.R., de Carvalho R.R., Odewahn S.C., Djorgovski S.G., Mahabal A.A., Brunner R.J., Lopes P.A.A. 2004, AJ, 128, 3082
\bibitem[]{} Gal R.R., Lopes P.A.A., de Carvalho R.R., Kohl-Moreira J.L., Capelato, H.V., Djorgovski S.G. 2009, AJ, 137, 2981
\bibitem[]{} Girardi, M., Giuricin, G., Mardirossian, F., Mezzetti, M., \& Boschin, W. 1998, ApJ, 505, 74
\bibitem[]{} Goto, T., Yagi, M., Tanaka, M. \& Okamura, S., 2004, MNRAS, 348, 515
\bibitem[]{} Goto, T., 2005, MNRAS, 359, 1415
\bibitem[]{} Haines, C.P., La Barbera, F., Mercurio, A., Merluzzi, P. \& Busarello, G., 2006, ApJ, 647, L21
\bibitem[]{} Hartigan, J. A. \&  Hartigan, P.M., 1985, The Annals of Statistics, 13, 70
\bibitem[]{} Hou, A., Parker, L., Harris, W. \& Wilman, D.J., 2009, ApJ, 702, 1199
\bibitem[]{} Hou, A., Parker, L. C., Wilman, D.J., McGee, S. L., Harris, W. E., Connelly, J. L., Balogh, M. L., Mulchaey, J. S. \& Bower, Richard G., 2012, MNRAS, 421, 3594
\bibitem[]{} Hoyle, F., Vogeleu, M.S. \& Pan, D., 2012, MNRAS, 426, 3041
\bibitem[]{} Kimm, T. et al., 2009, MNRAS, 394, 1131
\bibitem[]{} Katgert P. et al. 1996, A\&A, 310, 8
\bibitem[]{} Kauffmann, G. et al., 2004, MNRAS, 353, 713
\bibitem[]{} Krause, M.O., Ribeiro, A.L.B. \& Lopes, P.A.A., 2013, A\& A, 551, A143
\bibitem[]{} Lisker, T., Grebel, E.K. \& Binggeli, B., 2006, ApJ, 132, 497
\bibitem[]{} Lopes P.A.A. 2003,
Ph.D.\ Thesis, Observat\'orio Nacional (Brazil)
\bibitem[]{} Lopes P.A.A., de Carvalho R.R., Gal R.R., Djorgovski S.G., Odewahn S.C., Mahabal A.A., Brunner R.J. 2004, AJ, 128, 1017
\bibitem[]{} Lopes
  P.A.A., de Carvalho R.R., Capelato H.V., Gal R.R., Djorgovski S.G.,
  Brunner R.J., Odewahn S.C., Mahabal A.A. 2006, ApJ, 648, 209
\bibitem[]{} Lopes, P.A.A., 2007, MNRAS, 380, 1680
\bibitem[]{} Lopes P.A.A., de Carvalho R.R., Kohl-Moreira J.L., Jones C., 2009a, MNRAS, 392, 135 
\bibitem[]{} Lopes, P.A.A., de Carvalho, R.R., Kohl-Moreira, J.L., Jones, C., 2009b, MNRAS, 399, 2201
\bibitem[]{} Mann, A., \& Ebeling, H., 2012, MNRAS, 420, 2120
\bibitem[]{} Mart\'{\i}nez, H.J., Coenda, V. \& Muriel, H., 2008, MNRAS, 391, 585
\bibitem[]{} Mart\'{\i}nez, H.J. \& Zandivarez, A., 2011, MNRAS, 419, L24
\bibitem[]{} McGee, S.L., Balogh, M.L., Henderson, R.D.E., Wilman, D.J., Bower R.G.,
Mulchaey, J.S., Oemler, A., Jr., 2008, MNRAS, 387, 1605
\bibitem[]{} Menci, N. \& Fusco-Femiano, R., 1996, ApJ, 472, 46
\bibitem[]{} Mo, H., van den Bosch, F. \& White, S., 2010, in Galaxy Formation and Evolution, Cambridge University Press.
\bibitem[]{} Moore, B., Katz, N. \& Lake, G., 1998, ApJ, 457, 455
\bibitem[]{} Moss, C. \& Dickens, R.J., 1977, MNRAS, 178, 701
\bibitem[]{} Odewahn S.C., de Carvalho R.R., Gal R.R., Djorgovski S.G.,
Brunner R.J., Mahabal A.A., Lopes P.A.A., Kohl Moreira J.L.,
Stalder B., 2004, AJ, 128, 3092
\bibitem[]{} Merritt, D. 1987, ApJ, 313, 121
\bibitem[]{} Muirhead, J.R., 1982, Aspects of Multivariate  Statistical Theory, Wiley.
\bibitem[]{} Muratov, A.L. \& Gnedin, O.Y., 2010, ApJ, 718, 1266
\bibitem[]{} Navarro, J.F., Frenk, C.S. \& White, S.D.M., 1997, ApJ, 490, 493
\bibitem[]{} Ostriker J.P., 1980, Comments Astrophys., 8, 177
\bibitem[]{} Patiri,S.G., Prada, F., Holtzman, J., Klypin, A. \& Betancort-Rijo, J., 2006, MNRAS, 372, 1710
\bibitem[]{} Paudel, S., Lisker, T., Kuntschner, H., Grebel, E.K. \& Glatt, K., 2010, MNRAS, 405, 800
\bibitem[]{} Perea, J., del Olmo, A. \& Moles, M., 1990, A\& A, 237, 319
\bibitem[]{} Pinkney, J., Roettiger, K., Burns, J.O. \& Bird, C., 1996, ApJ, 104, 1
\bibitem[]{} Popesso P., Biviano A., B\"ohringer H., Romaniello M., 2006, A\&A, 445, 29
\bibitem[]{} Popesso, P., Biviano, A., Romaniello, M. \& B\"ohringer, H., 2007, A\& A, 461, 411
\bibitem[]{} Popesso, P., Biviano, A., B\"ohringer, H., Romaniello, M., 2007, A\&A, 464, 451
\bibitem[]{} Press, W.H., Teukolsky, S.A., Vetterling, W.T. \& Flannery, B.P., 1992,
Numerical Recipes in C (second ed.; Cambridge: Cambridge University Press)
\bibitem[]{} Ribeiro, A.L.B., Lopes, P.A.A \& Trevisan, M., 2010, MNRAS, 409, L124
\bibitem[]{} Ribeiro, A.L.B., Lopes, P.A.A. \& Trevisan, M., 2011, MNRAS, 413, L81
\bibitem[]{} Rines, K. \& Diaferio, A., 2006, AJ, 132, 1275
\bibitem[]{} Rood, H. J., Page, T. L., Kintner, E. C., \& King, I. 1972, ApJ, 175, 627
\bibitem[]{} S\'anchez-Bl\'azquez, P., Jablonka, P., Noll, S., Poggianti, B. M., Moustakas, J., Milvang-Jensen, B., Halliday, C., Aragón-Salamanca, A., Saglia, R. P., Desai, V., De Lucia, G., Clowe, D. I., Pell\'o, R., Rudnick, G., Simard, L., White, S. D. M. \& Zaritsky, D., 2009, A\& A, 499, 47
\bibitem[]{} S\'anchez-Janssen, R., Aguerri, J.A.L. \& Mu\~no\-Tu\~n\'on, C., 2008, ApJ, 679, L77
\bibitem[]{} Sehgal, N., et al., 2013, ApJ, 767, 38
\bibitem[]{} Sodr\'e, L.J., Capelato, H.V., Steiner, J.E. \& Mazure, A., 1989, AJ, 97, 1279
\bibitem[]{} Strateva I. et al., 2001, AJ, 122, 1861
\bibitem[]{} Team, R.D.C. (2011). R: A language and environment for statistical computing.
Vienna, Austria: R Foundation for Statistical Computiing. Retrieved from
http://www.R-project.org
\bibitem[]{} Thode, H.C., 2002, Testing for Normality, CRC Press
\bibitem[]{} Toloba, E., Boselli, A., Gorgas, J., Peletier, R. F., Cenarro, A. J., Gadotti, D. A., Gil de Paz, A., Pedraz, S. \& Yildiz, U., 2009, ApJ, 707, L17
\bibitem[]{} Trentham, N., 1998, MNRAS, 293, 71
\bibitem[]{} Wetzel, A.R., Tinker, J.L. \& Conroy, C., 2012, MNRAS, 424, 232
\bibitem[]{} White, M. Cohn, J. D. \& Smit, R., 2010, MNRAS, 408, 1818
\bibitem[]{} White, S. \& Rees, M., 1978, MNRAS, 183, 341
\bibitem[]{} Whitmore, B.C., Gilmore, D.M. \& Jones, C., 1993, ApJ, 407, 489
\bibitem[]{} Wolf, C., Gray, M.E. \& Meisenheimer, K., 2005, A\& A, 443, 435
\bibitem[]{} Wolf, C. et al., 2009, MNRAS, 393, 1302 
\bibitem[]{} Yahil, A. \& Vidal, N. V., 1977, ApJ, 214, 347
\bibitem[]{} Yee H. \& L\'opez-Cruz O., 1999, AJ, 117, 1985
\bibitem[]{} Yepes, G., Dominguez-Tenreiro, R., \& del Pozo-Sanz, R. 1991, ApJ, 373, 336
\bibitem[]{} Zabludoff, A.I. \& Mulchaey, J.S., 1998, ApJ, 496, 39
\bibitem[]{} Zitrin, A. et al., 2012, MNRAS, 426, 2944

\end{thebibliography}
\end{document}